\documentstyle[preprint,eqsecnum,floats,aps,epsf,tighten]{revtex}
\font\eightit=cmti8
\def\r#1{\ignorespaces $^{#1}$}

\newcommand{\met}      {\mbox{$\not\!\!{E}_{T} $}}

\begin{document}

\title{ Search for  Quark-Lepton Compositeness and a Heavy $W'$ Boson  
 Using the $e \nu$ Channel in $p \bar{p}$ Collisions at $\sqrt{s} = 1.8$ TeV}

\date{\today}
\maketitle

\font\eightit=cmti8
\def\r#1{\ignorespaces $^{#1}$}
\hfilneg
\begin{sloppypar}
\noindent
T.~Affolder,\r {23} H.~Akimoto,\r {45}
A.~Akopian,\r {37} M.~G.~Albrow,\r {11} P.~Amaral,\r 8  
D.~Amidei,\r {25} K.~Anikeev,\r {24} J.~Antos,\r 1 
G.~Apollinari,\r {11} T.~Arisawa,\r {45}
% ...
\end{sloppypar}
\vskip .026in
\begin{center}
(CDF Collaboration)
\end{center}

\vskip .026in
\begin{center}
\r 1  {\eightit Institute of Physics, Academia Sinica, Taipei, Taiwan 11529, 
Republic of China} \\
\r 2  {\eightit Argonne National Laboratory, Argonne, Illinois 60439} \\
\r 3  {\eightit Istituto Nazionale di Fisica Nucleare, University of Bologna,
I-40127 Bologna, Italy} \\
\r 4  {\eightit Brandeis University, Waltham, Massachusetts 02254} \\
\r 5  {\eightit University of California at Davis, Davis, California  95616} \\
\r 6  {\eightit University of California at Los Angeles, Los 
Angeles, California  90024} \\  
\r 7  {\eightit Instituto de Fisica de Cantabria, CSIC-University of Cantabria, 
39005 Santander, Spain} \\
\r 8  {\eightit Enrico Fermi Institute, University of Chicago, Chicago, 
Illinois 60637} \\
\r 9  {\eightit Joint Institute for Nuclear Research, RU-141980 Dubna, Russia}
\\
\r {10} {\eightit Duke University, Durham, North Carolina  27708} \\
\r {11} {\eightit Fermi National Accelerator Laboratory, Batavia, Illinois 
60510} \\
\r {12} {\eightit University of Florida, Gainesville, Florida  32611} \\
\r {13} {\eightit Laboratori Nazionali di Frascati, Istituto Nazionale di Fisica
               Nucleare, I-00044 Frascati, Italy} \\
\r {14} {\eightit University of Geneva, CH-1211 Geneva 4, Switzerland} \\
\r {15} {\eightit Glasgow University, Glasgow G12 8QQ, United Kingdom}\\
\r {16} {\eightit Harvard University, Cambridge, Massachusetts 02138} \\
\r {17} {\eightit Hiroshima University, Higashi-Hiroshima 724, Japan} \\
\r {18} {\eightit University of Illinois, Urbana, Illinois 61801} \\
\r {19} {\eightit The Johns Hopkins University, Baltimore, Maryland 21218} \\
\r {20} {\eightit Institut f\"{u}r Experimentelle Kernphysik, 
Universit\"{a}t Karlsruhe, 76128 Karlsruhe, Germany} \\
\r {21} {\eightit Center for High Energy Physics: Kyungpook National
University, Taegu 702-701; Seoul National University, Seoul 151-742; and
SungKyunKwan University, Suwon 440-746; Korea} \\
\r {22} {\eightit High Energy Accelerator Research Organization (KEK), Tsukuba, 
Ibaraki 305, Japan} \\
\r {23} {\eightit Ernest Orlando Lawrence Berkeley National Laboratory, 
Berkeley, California 94720} \\
\r {24} {\eightit Massachusetts Institute of Technology, Cambridge,
Massachusetts  02139} \\   
\r {25} {\eightit University of Michigan, Ann Arbor, Michigan 48109} \\
\r {26} {\eightit Michigan State University, East Lansing, Michigan  48824} \\
\r {27} {\eightit University of New Mexico, Albuquerque, New Mexico 87131} \\
\r {28} {\eightit The Ohio State University, Columbus, Ohio  43210} \\
\r {29} {\eightit Osaka City University, Osaka 588, Japan} \\
\r {30} {\eightit University of Oxford, Oxford OX1 3RH, United Kingdom} \\
\r {31} {\eightit Universita di Padova, Istituto Nazionale di Fisica 
          Nucleare, Sezione di Padova, I-35131 Padova, Italy} \\
\r {32} {\eightit University of Pennsylvania, Philadelphia, 
        Pennsylvania 19104} \\   
\r {33} {\eightit Istituto Nazionale di Fisica Nucleare, University and Scuola
               Normale Superiore of Pisa, I-56100 Pisa, Italy} \\
\r {34} {\eightit University of Pittsburgh, Pittsburgh, Pennsylvania 15260} \\
\r {35} {\eightit Purdue University, West Lafayette, Indiana 47907} \\
\r {36} {\eightit University of Rochester, Rochester, New York 14627} \\
\r {37} {\eightit Rockefeller University, New York, New York 10021} \\
\r {38} {\eightit Rutgers University, Piscataway, New Jersey 08855} \\
\r {39} {\eightit Texas A\&M University, College Station, Texas 77843} \\
\r {40} {\eightit Texas Tech University, Lubbock, Texas 79409} \\
\r {41} {\eightit Institute of Particle Physics, University of Toronto, Toronto
M5S 1A7, Canada} \\
\r {42} {\eightit Istituto Nazionale di Fisica Nucleare, University of Trieste/
Udine, Italy} \\
\r {43} {\eightit University of Tsukuba, Tsukuba, Ibaraki 305, Japan} \\
\r {44} {\eightit Tufts University, Medford, Massachusetts 02155} \\
\r {45} {\eightit Waseda University, Tokyo 169, Japan} \\
\r {46} {\eightit University of Wisconsin, Madison, Wisconsin 53706} \\
\r {47} {\eightit Yale University, New Haven, Connecticut 06520} \\
\r {(\ast)} {\eightit Now at Carnegie Mellon University, Pittsburgh,
Pennsylvania  15213} \\
\r {(\ast\ast)} {\eightit Now at Northwestern University, Evanston, Illinois 
60208}
\end{center}

\begin{abstract}
 We present searches for quark-lepton compositeness and a heavy $W'$ boson
 at high
 electron-neutrino transverse mass. We use $\sim$110
  pb$^{-1}$ of data collected in 
 {\mbox{$p\bar p$}}\ collisions at {\mbox{$\sqrt{s}$ =\ 1.8\ TeV }}
 by the CDF collaboration during 1992--95.  
 The data are consistent with standard model expectations. 
 Limits are set on the quark-lepton compositeness scale $\Lambda$ and
 the ratio of partial cross sections
 $\sigma (W' \rightarrow e \nu) / \sigma (W \rightarrow e \nu)$.
 The cross section ratio is used to obtain a lower limit 
 on the mass of a $W'$ boson with standard model couplings. We exclude 
 $\Lambda < 2.81$ TeV and a $W'$ boson with mass below 754 GeV/c$^2$
  at the 95\%
 confidence level. We combine the
 $W'$ mass limit with our previously published limit obtained using the 
 muon channel, to exclude a $W'$ boson with mass below 786 GeV/c$^2$ 
at the 95\% confidence level. 
\end{abstract}

\pacs{PACS numbers: 12.60.Rc, 13.85.Qk}

\narrowtext

 
 The standard model (SM) gives a good description of nature in terms of the
 fundamental fermions and their interactions via gauge bosons.
 However, the SM is not expected to be a complete 
 theory. For example,
  it does not explain the number of fermion families or their
 mass hierarchy. It also does not provide a unified description of all 
 gauge symmetries. 
 Compositeness models postulate constituents of the SM fermions and
 new strong dynamics that bind these constituents~\cite{llmodel}. 
 Other extensions of
 the SM postulate larger gauge groups and therefore new forces associated
 with  additional charged gauge bosons, which we generically call $W'$.
 For instance, the left-right symmetric 
 model~\cite{lrm}
 expands the $SU(2)_L \times U(1)$ electroweak group to
 $SU(2)_L \times SU(2)_R \times U(1)$, predicting an additional right-handed
 charged gauge boson.  

 At center-of-mass energies much smaller than the compositeness energy scale
 $\Lambda$,
 interactions between composite quarks and/or leptons have been parameterized
 by effective four-fermion contact interactions~\cite{llmodel}.  
 Atomic parity violation experiments have set stringent, though model-dependent
 limits on quark-lepton compositeness in the neutral current
 channel~\cite{parityvioexp}.
 Direct searches have set limits on $\Lambda$ in the
  range 2.5--6.1 TeV~\cite{d0comp,cdfcomp,opal} in a broad class
 of neutral current models. In this Letter, we present
 the first results of a search for compositeness in the charged current
 channel ($q \bar{q}' e \nu$) using the $e \nu$ final state. 

 The $e \nu$ final state is also sensitive to the direct production and decay
 of a $W'$ boson. 
 Previous indirect searches based on $\mu$ decay, the $K_L - K_S$
 mass difference, neutrinoless double beta decay,
 and studies of $b$ particles have resulted in stringent
 model-dependent limits on possible $W'$ bosons~\cite{langacker}.
  Direct searches in various decay modes have
 produced lower limits on the $W'$ mass, $m_{W'}$. The best limit of
 $m_{W'} > 720$ GeV/c$^2$
 in the $W' \rightarrow e \nu$ channel~\cite{d0wprime} 
 assumes a light and stable 
 neutrino, standard model couplings for the $W'$ to fermions,  
 and suppressed $W' \rightarrow WZ$ decays, as in extended gauge 
 models~\cite{altarelli}. In this Letter,  we set upper limits on 
 the ratio of partial cross sections
 $\sigma (W' \rightarrow e \nu) / \sigma (W \rightarrow e \nu)$  under
 the same assumptions. We use the latter to obtain the most stringent
  lower limit on $m_{W'}$. We also present the combined 
 $W'$ mass limit with our previously published limit obtained using the 
 muon channel~\cite{wprimemu}. 

 
 We use $\sim$110 pb$^{-1}$ of data collected in 
 $p\bar{p}$ collisions at  $\sqrt{s}=1.8$ TeV by the  Collider Detector
 at Fermilab 
 \cite{cdfdetector} during 1992--95.
 The  detector includes a tracking system immersed in a 1.4 T magnetic
 field, scintillator-based sampling electromagnetic and hadronic calorimeters,
 and a muon detector.  
 For this analysis, electron candidates are accepted in the 
 pseudorapidity range $0.05 <$ $\mid$$\eta$$\mid$ $<1.0$, 
 where $\eta = -{\rm log \; tan} (\theta/2)$, and $\theta$
 is the polar angle with respect to the beam axis.
 Electrons detected near the fiducial edges of the calorimeter are removed
 to ensure uniform calorimeter response. 
 We use a combination of electron and 
 neutrino triggers to obtain an 
 efficiency exceeding 99\%  for the high transverse mass $e \nu$ final
 states that pass our offline selection criteria. 

  After offline reconstruction, the electromagnetic calorimeter cluster  
 with the highest transverse energy ($E_T \equiv E \, {\rm sin} \, \theta$) 
 in the event 
 must satisfy these requirements: 
 (i) the electron must deposit most~\cite{hadem} of its energy in the 
     electromagnetic calorimeter, 
 (ii) a track in the central drift chamber must match the calorimeter 
 cluster in position, 
 (iii) the electron must be isolated in a cone of radius 
 $R \equiv \sqrt{\Delta \eta^2 + \Delta \phi^2} = 0.4$, 
 such that the fractional excess transverse
 energy in the cone, $\frac{E_T^{tot}(R=0.4)
 - E_T^e}{E_T^e}<0.1$, where $E_T^{tot}$ and $E_T^e$ are the
 total and electron transverse energies respectively. 
 The kinematic cuts used to define the data sample are $E_T^e > 30 $ GeV, 
 the transverse momentum ($p_T$) of the associated track $p_T^e >
 $ 13 GeV/c, the missing transverse energy \met\ $ > 30$ GeV, and the
 electron-neutrino transverse mass $m_T (e \nu) >$ 50 GeV/c$^2$, where 
 $m_T ( e \nu) = 
    \sqrt{2 \; E_T^e \; \met \; (1 - \cos \; \phi_{e \nu }) }$, and
 $\phi_{e \nu}$ is the azimuthal angle between the electron and the \met\ 
 direction. The neutrino
 transverse momentum is identified with \met\ by requiring 
 transverse momentum balance in the event. Electron 
 identification cuts based on $E/p$ (ratio of 
 calorimeter energy to matched track momentum) and calorimeter energy
 profiles, which are imposed for $E_T^e < 50$ GeV to suppress jet
 misidentification backgrounds, are released for $E_T^e > 50$ GeV to ensure
 maximum signal efficiency. A total of
 31,436 events pass our selection criteria. 

 We use  the 
 {\small PYTHIA}~\cite{pythia} program
 to compute the compositeness and $W'$ signal processes.
 The detector response is simulated using a parameterized
 Monte Carlo program. The electromagnetic calorimeter
 sampling term is  derived  from
 test beam data. The underlying event contribution
  to the electron energy resolution is derived from 
 $W \rightarrow e \nu$ collider data. 
 The constant term in the electromagnetic resolution
 is tuned to reproduce the observed width 
 of the  $Z \rightarrow ee$ mass peak. 
 The electromagnetic energy
 scale is set so that the reconstructed $Z$ boson mass agrees with the 
 world-average 
 $Z$ mass~\cite{bayes}.
 The hadronic response and resolution are tuned by studying the
 $p_T$ balance in $Z \rightarrow ee$ events. 

 In this analysis 
 we normalize the number of SM background 
 Monte Carlo events after detector simulation to the large
 inclusive $W$ boson sample in the data. Thus we are analysing the shape of
 the $e \nu$ transverse mass distribution, and are insensitive to the
 uncertainty in the integrated luminosity of the data and to the overall
 efficiency. 
 The efficiency of the additional electron identification cuts applied
 for $E_T^e < 50$ GeV is determined 
 using $ Z \rightarrow ee$ data where one of the electrons is tagged.
  The second electron then 
 provides an unbiased sample 
 with which
 to measure the efficiencies. Background subtraction is performed using the
 sidebands of the $Z$ boson mass distribution.
 The combined efficiency of these additional cuts
  is (95.8$\pm$0.3)\%, relative to the full efficiency at high  $E_T^e$.

 The most important sources of misidentification
 background to $ p \bar{p} \rightarrow e \nu + X$ 
 are (i) QCD multijet events, where a jet is misidentified as an electron and
 there is sufficient energy mismeasurement to create significant \met,  and 
 (ii) $Z \rightarrow ee$ events where one electron is
  lost or misreconstructed. 
 The electromagnetic energy in a jet which has been misidentified as an
 electron is likely to be non-isolated. We select a representative 
 sample of misidentified
 electrons by making the electron identification cuts on the base
 sample without the isolation cut, and then selecting non-isolated candidates. 
 The relative normalization of this sample to the jet background in the
 signal sample is obtained from a ``pure-jet'' sample. The  ``pure-jet''
 sample is obtained in the same way as the signal sample except \met $< 10$
 GeV, which excludes almost all $W$ events. This technique assumes that the 
 isolation for a jet 
 is independent of \met. The systematic uncertainty of 30\% on
 the jet misidentication background is estimated by studying the 
 correlation between isolation and \met. The $Z \rightarrow ee$ background
 is estimated using a Monte Carlo sample of $Z \rightarrow ee$ events, passed
 through a full detector simulation and reconstructed like the data. The 
 sytematic uncertainty of 23\% on the $Z \rightarrow ee$ background is 
 estimated by varying the detector response to electrons near the fiducial 
 edges of the calorimeter. Other systematic uncertainties, indicated in 
 Table~\ref{systematicerror}, are derived by varying the parameters in the
 Monte Carlo simulation. The uncertainty due to parton distribution functions
 is taken to be identical with our published $W' \rightarrow \mu \nu$ 
 analysis~\cite{wprimemu}. 
  
\begin{table}[hbtp]
\medskip
\caption{The observed number of events and the total expected number
 of events from SM and detector background sources, in transverse
 mass bins.}
\medskip
\begin{tabular}{|l|c|c|}
  $m_T$ bin (GeV/c$^2$)  & $N_{\rm observed}$ & $N_{\rm expected}$ \\
\hline
\hline
150-200 & 70 &  62.2$\pm$8.5  \\
\hline
200-250 & 18 &  18.3$\pm$3.4  \\
\hline
250-300 & 5 &  4.01$\pm$0.44  \\
\hline
300-350 & 2 &  1.61$\pm$0.18  \\
\hline
350-400 & 0 &  0.72$\pm$0.08  \\
\hline
400-500 & 1 &  0.49$\pm$0.06  \\
\hline
500-600 & 0 &  0.11$\pm$0.02  \\
\hline
600-1000 & 0 &  0.05$\pm$0.01  \\
\end{tabular}
\label{ndatamc}
\end{table}

 Other
 high $p_T$  processes  also contribute to $e \nu$ final states.
 Using {\small PYTHIA}, we evaluate the following background
 processes, $W \rightarrow e  \nu$ (dominant), 
 $W \rightarrow \tau  \nu \rightarrow e \nu X$, 
 $ t\bar{t} \rightarrow e \nu X$, $WW \rightarrow e \nu X$, 
 $WZ \rightarrow e \nu X$, $ZZ \rightarrow e \nu X$ and 
 $ \gamma^*/Z \rightarrow \tau \tau \rightarrow e \nu X $.
 We pass these Monte Carlo events through the
 parameterized detector simulation to estimate their contribution. These 
 physics backgrounds dominate over the misidentification backgrounds at high
 $e \nu$ transverse mass, due to the presence of real neutrino(s) producing
 large \met. For example, the jet and $Z \rightarrow ee$ misidentification 
 background fractions amount to 25\% and 3\% respectively for 
 $m_T (e \nu) > 150$ GeV/c$^2$. 

\begin{table}[hbtp]
\caption{Systematic uncertainties on the SM background and the signal
  due to  the parton distribution functions (PDFs), 
the $K$-factor, and  the detector model.
}
\medskip
\begin{tabular}{|l|c|c|}
                    & SM Background (\%) & Signal (\%) \\
\hline
\hline
PDFs & 10 & 10 \\
\hline
$K$-factor & 4 & 4 \\
\hline
\hline
hadronic resolution & 0.1 & 2\\
\hline
vertex $z$ width    & 0.5   & 1.8 \\
\hline
hadronic scale & 0.2 & 1.6\\
\hline
EM resolution  & 0.1 & 1.5  \\
\hline
electron efficiency    & 1.0   & 1.0 \\
\hline
EM scale & 0.2 &  0.9  \\
\hline
\hline
total & 11 & 12 \\
\end{tabular}
\label{systematicerror}
\end{table}

\begin{figure}[tbhp]
\epsfxsize=3.0in
\centerline{\epsfbox{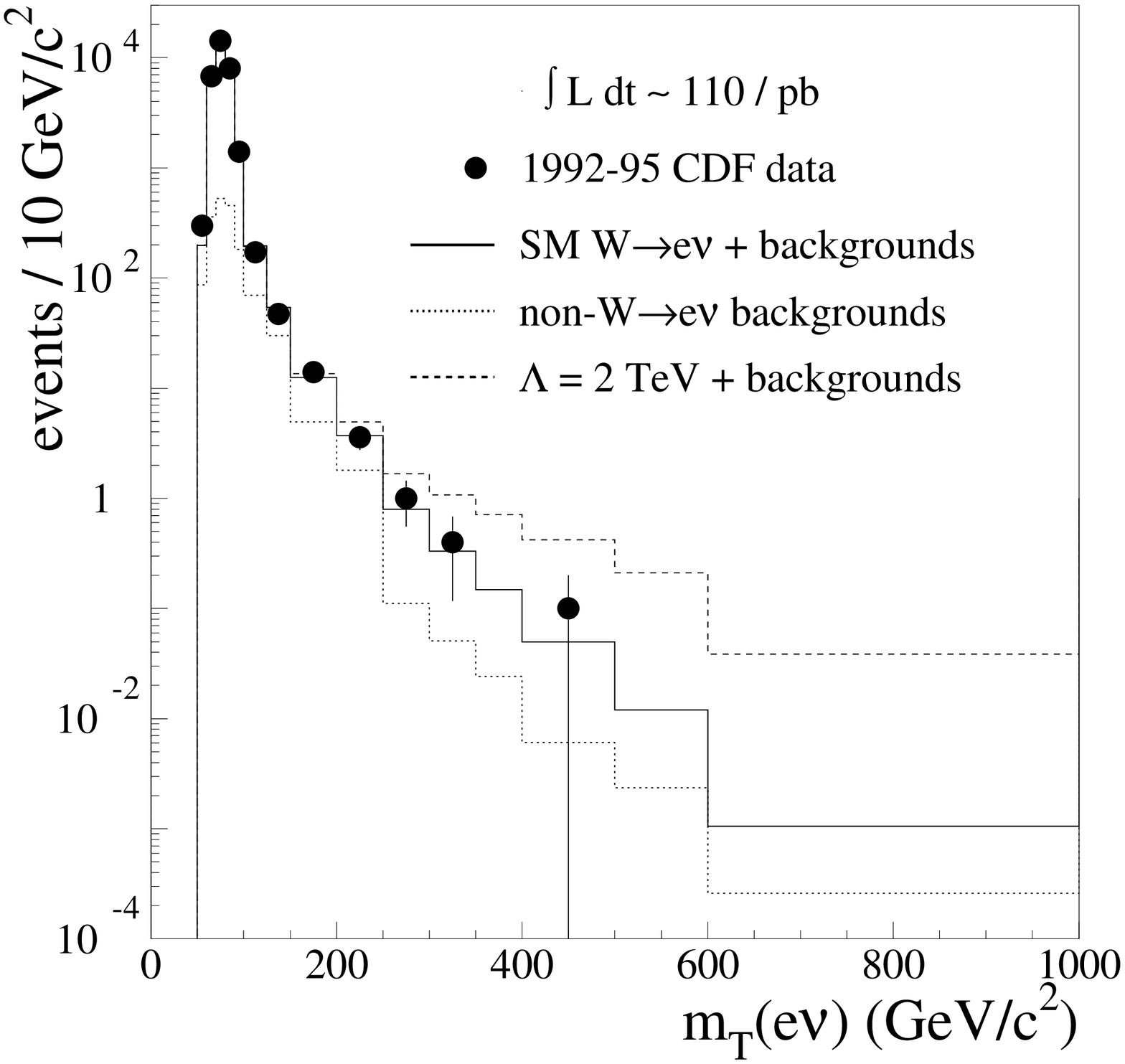}}
\caption {The event yield from the data as a function of the $e \nu$ transverse
 mass, normalized to the bin width. Also shown are the SM prediction including
 backgrounds, all backgrounds excluding the dominant SM $W \rightarrow e \nu$
 process, and the prediction of the compositeness process with energy scale
 $\Lambda = 2$ TeV. The simulation of the physics processes includes the 
 effects of detector acceptance and response. 
}
\label{wmt_dt4}
\end{figure}

 Figure~\ref{wmt_dt4} shows the transverse mass distribution of the data
 events normalized to the bin width. Also shown is the expectation based on
 SM processes and detector backgrounds. 
 We apply a mass-dependent
 $K$-factor (defined as the ratio of the next-to-next-to leading order (NNLO)
  and the leading-order (LO)
  Drell-Yan cross section calculations from Ref.~\cite{dynnlo}) to the LO
  {\small PYTHIA} calculation. The $K$-factor varies between 1.24 at 
 80 GeV/c$^2$
 and 1.65 at 800 GeV/c$^2$.  
 The effects of the detector
 acceptance and response have been folded into the theoretical prediction.
 Table~\ref{ndatamc} shows the expected and the observed number of events
 in the high transverse mass bins. 
 There is good agreement between the
 data and the expectation. Also  shown are all backgrounds excluding the 
 dominant SM $W \rightarrow e \nu$ process, and 
 the expectation of the compositeness
 model with $\Lambda = 2$ TeV. 

 To set a limit on the compositeness scale $\Lambda$, 
 we generate Monte Carlo events for the compositeness process 
 using {\small PYTHIA}, corrected with the $K$-factor. 
 We perform a 
 Bayesian analysis~\cite{bayes}
  of the shape of the $m_T$ distribution of events.
 The expected number of events in the 
 $k^{\rm th}$ transverse mass bin is denoted by 
 $N_{\Lambda}^k = b^k + {\cal L}{\epsilon^k} {\sigma_{\Lambda}^k}$, 
 where  ${\sigma_{\Lambda}^k}$
 is the predicted cross section for a given scale $\Lambda$, 
 and $\epsilon^k$ and $b^k$ denote the total acceptance and remaining 
 backgrounds in the $k^{\rm th}$ bin. The prediction for the number of events,
 including all backgrounds, is normalized to the observed number of events for
 $m_T (e \nu) < 150$ GeV/c$^2$. 
 Given the data $(D)$, we compute the posterior probability distribution
 for $\Lambda$ according to
\begin{eqnarray}
P({\Lambda}{\mid}{D})=
\frac{1}{A}
\int db \; {d\epsilon}
\prod_{k=1}^n 
\left[ 
\frac {{e^{-{N_{\Lambda}^k}}}{N_{\Lambda}^k}^{N_{o}^k}} {{N_{o}^k}!}
P({b^k},
{\epsilon^k})
\right] 
P({\Lambda}). \nonumber
\label{post_prob}
\end{eqnarray} 
 $N_{o}^k$ denotes the observed number of events. 
 We take 
 the prior distribution $P({b^k},{\epsilon^k})$ of the nuisance parameters 
 $b$ and $\epsilon$ to be Gaussian with the r.m.s. given by their
 total uncertainties. The bin-to-bin correlations in 
  the uncertainty on the acceptance and 
 background  are taken into account.  
 We make the conventional choice for the
  prior distribution $P({\Lambda})$ to be  
 uniform in 1/${\Lambda^2}$. 
 The 95\% C.L. lower limit
  is defined by
  $ \int_{\Lambda}^{\infty} P({\Lambda^\prime}{\mid}{D}) d{\Lambda^\prime}
 =0.95$, yielding $\Lambda > 2.81$ TeV. The expected limit, obtained when
 the observed number of events is set equal to the expected number, is 
 $\Lambda > 2.70$ TeV. Varying the choice of the prior distribution 
 $P({\Lambda})$ changes the limit by 10\%. 

\begin{figure}[tbhp]
\epsfxsize=3.0in
\centerline{\epsfbox{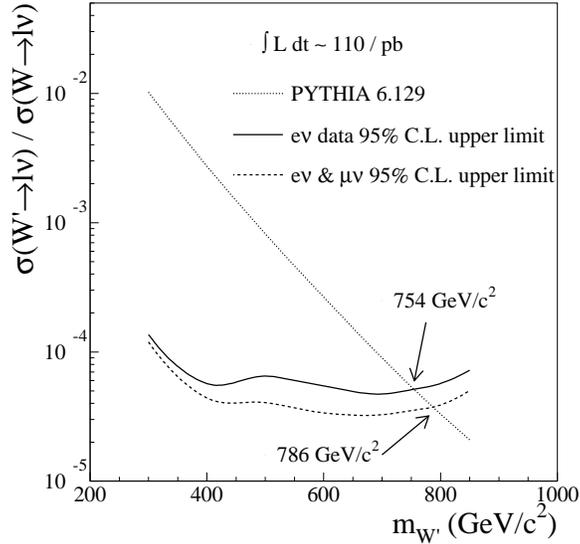}}
\caption {The 95\% C.L. upper limit on the ratio of partial cross sections
 $\sigma (W' \rightarrow \ell \nu) / \sigma (W \rightarrow \ell \nu)$, for the
 $e$ data and the combined $e + \mu$ data. Also shown
 is the SM prediction for this ratio, and the $m_{W'}$ limits obtained from
 the intersection of the experimental and theory curves. 
}
\label{wmt_shape_wprime_plot4}
\end{figure}

 To set a limit on the mass of a $W'$ boson, we 
 compute the Poisson probability for the
 observed number of events given $N_{\rm expected} = N_{\rm background} + 
 N_{W'}$. The Poisson probability is computed separately in three search
 windows: $0.5 M_{W'} < m_T < 0.65 M_{W'}$, 
 $0.65 M_{W'} < m_T < 0.8 M_{W'}$ and
 $0.8 M_{W'} < m_T < 1.1 M_{W'}$, and then the probabilities are combined. 
 The use of three windows allows us to exploit the difference in the shape
 of the $W'$ signal and background $m_T$ distributions. Uncertainties in
 the backgrounds and signal acceptance are incorporated by convoluting
 the probability $P(N_{W'})$
 over Gaussian fluctuations in these parameters, taking
 correlations across bins into account. The 95\% C.L. upper limit on
 the number of $W'$ signal events, $N_{W'}^{95}$, is  defined by 
$ \int_0^{N_{W'}^{95}} P(N_{W'}) d(N_{W'}) = 0.95 $. 
The limit $N_{W'}^{95}$ may be expressed as a 95\% C.L. limit on the ratio
 $\sigma B (W' \rightarrow e \nu) / \sigma B (W \rightarrow e \nu)$ using
\begin{eqnarray}
\left( \frac{\sigma B (W' \rightarrow e \nu)}
     { \sigma B (W \rightarrow e \nu)} \right) _{95} = 
\frac{ N_{W'}^{95} A_W }{ A_{W'} N_W} \nonumber
\end{eqnarray}
 where $N_W$ is the observed number of SM $W$ events and $A_{W'}$($A_{W}$)
 is the total 
 acceptance for $W' \rightarrow e \nu$ ($W \rightarrow e \nu$) decays. 
 The 95\% C.L. upper limit on 
 $\sigma B (W' \rightarrow e \nu) / \sigma B (W \rightarrow e \nu)$ is
 plotted as a function of $M_{W'}$ in Fig.~\ref{wmt_shape_wprime_plot4}
  together with the theory
 curve from {\small PYTHIA} 6.129, assuming standard model couplings
 and including the $K$-factor. 
 From the intersection of the two curves, a $W'$ boson with mass
 $m_{W'} < 754$ GeV/c$^2$
 is excluded at 95\% C.L. The expected limit in this case is
 748 GeV/c$^2$. We combine this result with our previously published
 result on a $W'$ boson using the $\mu \nu$ final state~\cite{wprimemu}. 
 Taking the PDF uncertainty to be fully correlated between the two 
 analyses and with the same model assumptions, we obtain the combined limit
 excluding $m_{W'} < 786$ GeV/c$^2$ at the 95\% C.L. 

 In conclusion, we find no significant deviation between
  the measured $ e \nu$ transverse
 mass distribution at high transverse mass and the SM prediction. 
 We have used the data to exclude  the
 quark-lepton compositeness scale $\Lambda < 2.81$ TeV, 
 in the context of an effective 
 contact interaction. We set limits on the ratio of the cross section times
 branching ratio to $ e \nu$ of a $W'$ boson to a standard model $W$ boson. 
 We use the latter to exclude a $W'$ boson with SM couplings and mass 
 $m_{W'} < 754$ GeV/c$^2$. Combining with our muon channel result, we exclude 
 $m_{W'} < 786$ GeV/c$^2$ at the 95\% C.L. 
    
We thank T. Sj$\ddot{\rm o}$strand for discussions 
regarding {\small PYTHIA} and W. L. Van Neerven for providing the code 
 to compute the 
NNLO SM Drell-Yan cross section. 
          We thank the Fermilab staff and the technical staffs of the
participating institutions for their vital contributions.  This work was
supported by the U.S. Department of Energy and National Science Foundation;
the Italian Istituto Nazionale di Fisica Nucleare; the Ministry of Education,
Science, Sports and Culture of Japan; the Natural Sciences and Engineering 
Research Council of Canada; the National Science Council of the Republic of 
China; the Swiss National Science Foundation; the A. P. Sloan Foundation; the
Bundesministerium fuer Bildung und Forschung, Germany; the Korea Science 
and Engineering Foundation (KoSEF); the Korea Research Foundation; and the 
Comision Interministerial de Ciencia y Tecnologia, Spain.

\begin{references}
\bibitem{hadem} For electron candidates we require $E_{had}/E_{em} < 0.055 +  
 0.045 \times (\; E^e \; / \; 100 \; {\rm GeV} \; )$, 
 where $E_{had}$ and $E_{em}$ are
 the hadronic and electromagnetic energies respectively.
\end{references}
\end{document}